# Graphene/CdTe heterostructure solar cell and its enhancement with photo-induced doping

Shisheng Lin,*†‡ Xiaoqiang Li, † Shengjiao Zhang, † Peng Wang, † Zhijuan Xu, † Huikai Zhong, † Zhiqian Wu, †Hongsheng Chen

†Department of Information Science and Electronic Engineering, Zhejiang University, Hangzhou, 310027, China

‡ State Key Laboratory of Modern Optical Instrumentation, Zhejiang University, Hangzhou, 310027, China

*Corresponding author. Email: shishenglin@zju.edu.cn

**We report a type of solar cells based on graphene/CdTe Schottky heterostructure, which can be improved by surface engineering as graphene is one-atomic thin. By coating a layer of ultrathin CdSe quantum dots onto graphene/CdTe heterostructure, the power conversion efficiency is increased from 2.08% to 3.1%. Photo-induced doping is mainly accounted for this enhancement, as evidenced by transport, photoluminescence and quantum efficiency measurements. This work demonstrates a feasible way of designing solar cells with incorporating one dimensional and two dimensional materials.**

Graphene possesses several outstanding properties including high intrinsic carrier mobility, [1] micro-scale ballistic transport, [2] abnormal quantum Hall effect, [3] 97.7% constant transmittance of visible light, [4] extraordinary thermal conductivity [5] and high

Young's modulus.[6] These unique properties make graphene an outstanding material for optoelectronic applications.[7, 8] Besides, the Fermi level of graphene is tunable as the density of states near the Dirac point is low. The richness of optical and electronic properties of graphene makes itself a great candidate for optoelectronic devices, where doping of graphene is essential for high performance. It has been well reported that electrical properties of graphene can be tuned by surface adsorption of foreign molecules,[9, 10] electrical gating[11, 12] and magnetic effects.[13, 14] Recently, there is a rising interest in photo-induced doping of graphene, which can be used to design various types of two-dimensional optoelectronic devices.[15-17]

Photo-induced doping represents a physical picture of photo-induced carrier injection, which could utilize the atomic sharp nature of graphene/bulk semiconductor heterostructure and low density of electronic states of graphene, although no work has been reported on the photo-induced doping effect in graphene based solar cells. Current attentions are mainly devoted to graphene/Si heterostructures.[18-20] Compared with Si, CdTe has a direct band gap of 1.45 eV, close to the ideal band gap for solar energy conversion.[21, 22] Meanwhile, CdTe based thin-film solar cell is one of the most important photovoltaic devices for cost-effective and clean generation of solar electricity,[23, 24] which is also suitable for flexible applications. Graphene/CdTe heterostructure should be one of the best choices for graphene based solar cells. However, to the best of our knowledge, no graphene/CdTe heterostructure solar cell has been reported. Moreover, as an interface solar cell, graphene/semiconductor heterostructure should receive more sophisticated design. Herein, a physical picture is illustrated by incorporating photo-induced doping into graphene/CdTe heterostructure solar cells by simply covering graphene with CdSe quantum dots (QDs). With photo-induced doping, the power conversion efficiency (PCE) is improved by 50%. This work presents a way for enhancing the performance of grahene/semiconductor heterostructure, which should be effective for graphene/Si and other graphene/semiconductor system as well.

CdTe with a thickness of 8μm was grown by vapor pressure deposition technique on polished glass covered with $SnO_2$:F conducting film. The grown CdTe film is p-type doped with hole concentration around $1\times10^{16}$ cm$^{-3}$. $MgCl_2$ activation treatment [24] at 420 °C for 20 min was done followed by nitric-phosphoric acid etching [25] before graphene transferring. Graphene was grown on copper substrate by chemical vapor deposition (CVD) technique using $CH_4$ and $H_2$ ($CH_4$:$H_2$ equals to 5:1) as the reaction gas sources at 1000 °C for 30 min. The active area (10 mm×1 mm) was defined by plasma enhanced CVD deposited $SiN_x$ (80 nm) with lithography-processed mask. Graphene was transferred onto the CdTe substrate using polymethyl methacrylate (PMMA) as the sacrificing layer. [26] After removal of PMMA layer with 400 °C anneal under $H_2$/Ar mixture atmosphere, Ag was pasted onto graphene above $SiN_x$ film and on $SnO_2$:F layer, followed by a post anneal at 120 °C for 5 min. For preparation of CdSe QDs, 0.0142 g dimethyl cadmium and 0.079 g selenium powder were co-dissolved in 50mL tri-alkyl phosphine, and then the solution was injected into 50 mL trioctylphosphine oxide (340–360 °C). After 10 seconds of nucleation, temperature was set at 280 °C for 2 min followed by cooling of the system. Dispersion solution of CdSe QDs in toluene (10 mg/mL) was spun onto graphene followed with 150°C/5min baking process. The graphene/CdTe solar devices were tested under AM1.5G condition. The illumination intensity was calibrated with a standard Si solar cell. The current density-voltage (J-V) data were recorded using Agilent B1500A system. Graphene was characterized by Raman spectroscopy (Horiba Jobin-Yvon LabRAM HR800 UV-vis μ-Raman and Renishaw inVia Reflex) with the excitation wavelength of 514 nm and 780 nm. Quantum efficiencies (QE) were measured with PV Measurements QEXL system. Absorbance of the CdSe QDs was measured with LabTech UV-Vis spectrophotometer. The microstructure of CdSe QDs on graphene was measured with FEI Tecnai F20. Photoluminescence (PL) properties of CdSe QDs were measured with a FLS920 fluorescence spectrometer (Edinburgh Instruments) at room temperature using a line of 405 nm as the excitation light.

Raman and resistance measurements are designed to confirm CdSe QDs introduced photo-induced doping in graphene. The set-up of the experiments is schematically shown in Figure 1(a). Back Au gate electrode is used as gate electrode to form field effect transistor (FET) structure. Graphene is usually get p-type doped after wet transferring.[27] The conductivity type of as-grown graphene after transferring in this work is measured by the resistance variation of graphene with different gate voltage. As shown in Figure 1(b), the peak value of resistance corresponds to gate voltage of +20 V, clarifying as-grown graphene is p-type doped. Figure 1(c) shows the Raman spectra of graphene with and without QDs using 514 nm laser as the excitation source, which has a spot size around 10 μm$^2$. For graphene without QDs, G peak locates at 1590 cm$^{-1}$, which is blue-shifted compared with 1580 cm$^{-1}$ for undoped graphene. The corresponding Fermi level position of graphene ($E_{F-Gra}$) can be deduced as 0.24 eV below Dirac point.[12] For graphene covered with QDs, when measured with 0.13mW laser power, G peak red-shifts to 1583cm$^{-1}$. The corresponding $E_{F-Gra}$ is 0.14 eV below or above Dirac point, which means that laser generated electrons in CdSe QDs have been injected into graphene and move $E_{F-Gra}$ upward during Raman measurements. The case for the Raman spectrum measured with 1.3 mW laser power is that G peak blue-shifts to the position 1601 cm$^{-1}$, which indicates that conduction type of graphene has been changed into n-type as high concentration of electrons injected from CdSe QDs and corresponds to Fermi level position of 0.50 eV above Dirac point. Figure 1(d) shows the transport behavior of graphene with or without QDs coverage between two lateral electrodes under xenon lamp illumination with different light intensity. For graphene without QDs, resistance stays constant. While for the case with QDs, resistance increases at low light intensity, the peak value shows up at light intensity of 2 suns. Then resistance of graphene decreases as light intensity increases. The turnover of the graphene resistivity indicates the transformation of conductivity type. As mentioned above, the as-grown graphene is p-type doped, the increase of the resistance attributes to electrons injected from CdSe QDs based on photo-induced doping. After the peak value, conduction type of graphene is changed into n-type, stronger photo-induced doping leads to lower resistance.

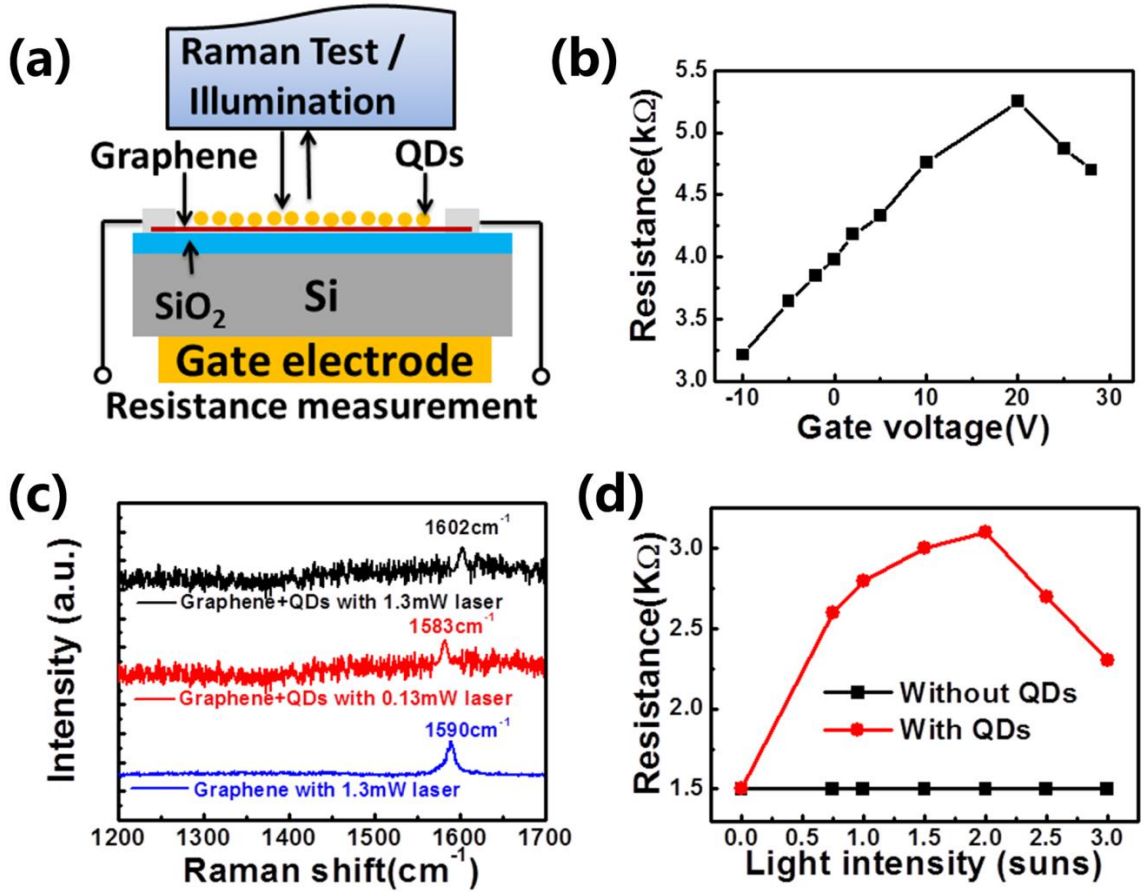

**Fig. 1.** (a) Setup of the Raman and resistance measurements. (b) Resistance of graphene under different gate voltage. (c) Raman spectra of graphene with and without QDs under laser illumination with different power. (d) Resistance variation of graphene under illumination with different light intensity.

Figure 2(a) shows the PL spectra of CdSe QDs with 405 nm excitation laser, where the PL intensity of CdSe QDs is one magnitude lower when the QDs is covering on graphene than that of the case without graphene underlying. PL quenching indicates that large part of the laser generated electrons have injected into graphene. Besides, PL spectra cover the wavelength range 450 nm-850 nm. The PL peak locates at 530 nm wavelength corresponds to the band-to-band recombination (2.34eV), and the wide PL band locates from 550 nm to 850 nm attributes to defects levels. Figure 2(b) schematically illustrates the mechanism of CdSe QDs introduced photo-induced doping. The photo-generated electrons from valence band and defects levels in CdSe QDs inject into graphene, while holes are trapped in the QDs. The injected electrons and positive charged QDs can greatly influence the carrier concentration in graphene.[17]

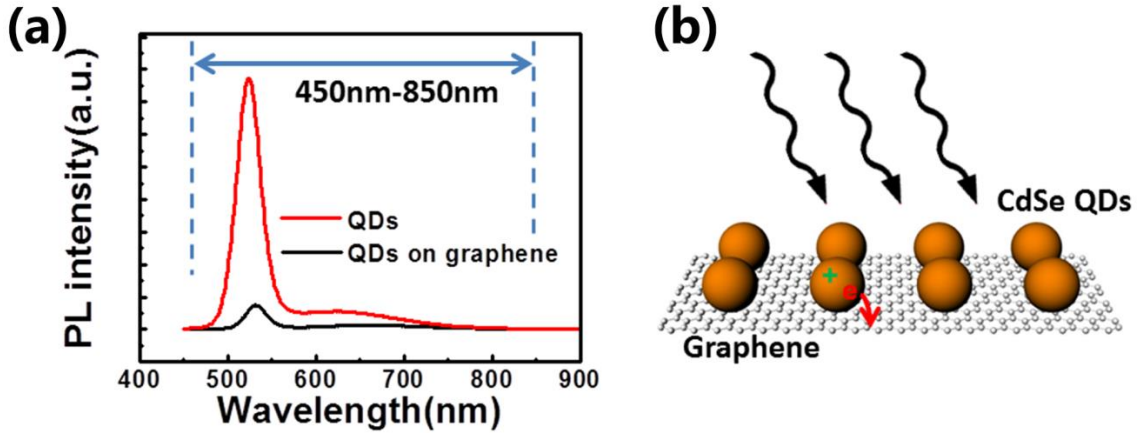

**Fig. 2.** (a) PL spectra of CdSe QDs with or without touching with graphene using the excitation laser line of 405 nm. PL is quenched for the CdSe QDs on graphene. (b) Mechanism of the CdSe QDs introduced photo-induced doping in graphene.

Photo-induced doping can be utilized in optoelectronic devices, in this study, we design photo-induced doping enhanced graphene/CdTe solar cells, the schematic structure is shown in Figure 3(a), which is composed of glass substrate, $SnO_2$:F conducting layer, p-type CdTe film, $SiN_x$ inter dielectric layer, graphene and Ag electrodes. CdSe QDs are placed on the surface of graphene. The devices are illuminated from the graphene side during measurements. Figure 3(b) shows the high resolution transmittance electron microscopy (HRTEM) image of the CdSe QDs. These QDs are single crystals with diameter of 2.4 nm-2.8 nm. The inset in Figure 3(b) shows the Fast Fourier Transform (FFT) image of the marked QD with red arrow. The measured interplane spacing is 0.35 nm, which matches well with the (0002) plane of CdSe.

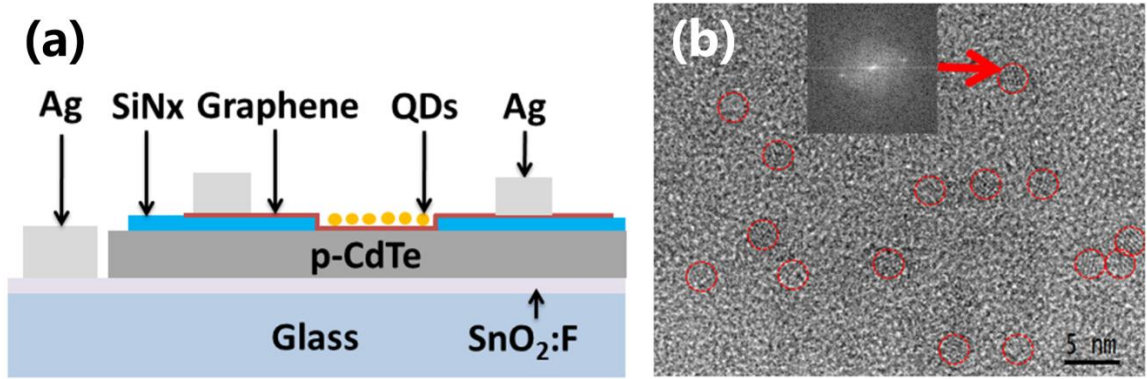

**Fig. 3.** (a) Schematic structure of the CdSe QDs covered graphene/CdTe solar cell. (b) HRTEM micrograph of the CdSe QDs.

J-V curves of the graphene/CdTe solar cells with and without QDs under AM1.5G illumination and in the dark are shown in Figure 4(a) and 4(b), respectively. Under illumination, , PCE is 2.18% for the device without QDs with the values of open circuit voltage ($V_{oc}$), short circuit current density ($J_{sc}$) and fill factor (FF) are 0.45 V, 13.5 mA/cm$^2$ and 34.2% respectively. After coating CdSe QDs on graphene, $J_{sc}$ increases to 16.4 mA/cm$^2$. Meanwhile, J-V curve down-shifts as $J_{sc}$ increases. $V_{oc}$ is increased to 0.51 V and FF is slightly increased to 37.1%. The final PCE for the graphene/CdTe solar cell with QDs is 3.10%. Dark J-V curves of the devices with and without QDs show similar rectifying characteristics, suggesting there is little doping effect from the covering CdSe QDs in the dark condition. External quantum efficiency (EQE) curves of the devices with and without QDs are plotted in Figure 4(c). For wavelengths below 400 nm, device without QDs has higher EQE, while for wavelengths above 400 nm, the result is opposite. Besides, there is a valley around 520 nm in the EQE curve for the device with QDs. Figure 4(d) shows the absorbance curve of the CdSe QDs. Absorbance at short wavelengths is very strong and there is an absorbance peak around 520 nm, which accounts for the low EQE at short wavelengths and around 520 nm for the device with

QDs. The size of the CdSe QDs can be estimated by the first absorption peak. [28] The deduced size of CdSe QDs corresponding to absorption peak at 520 nm is about 2.6 nm, which agrees well with the HRTEM measured results.

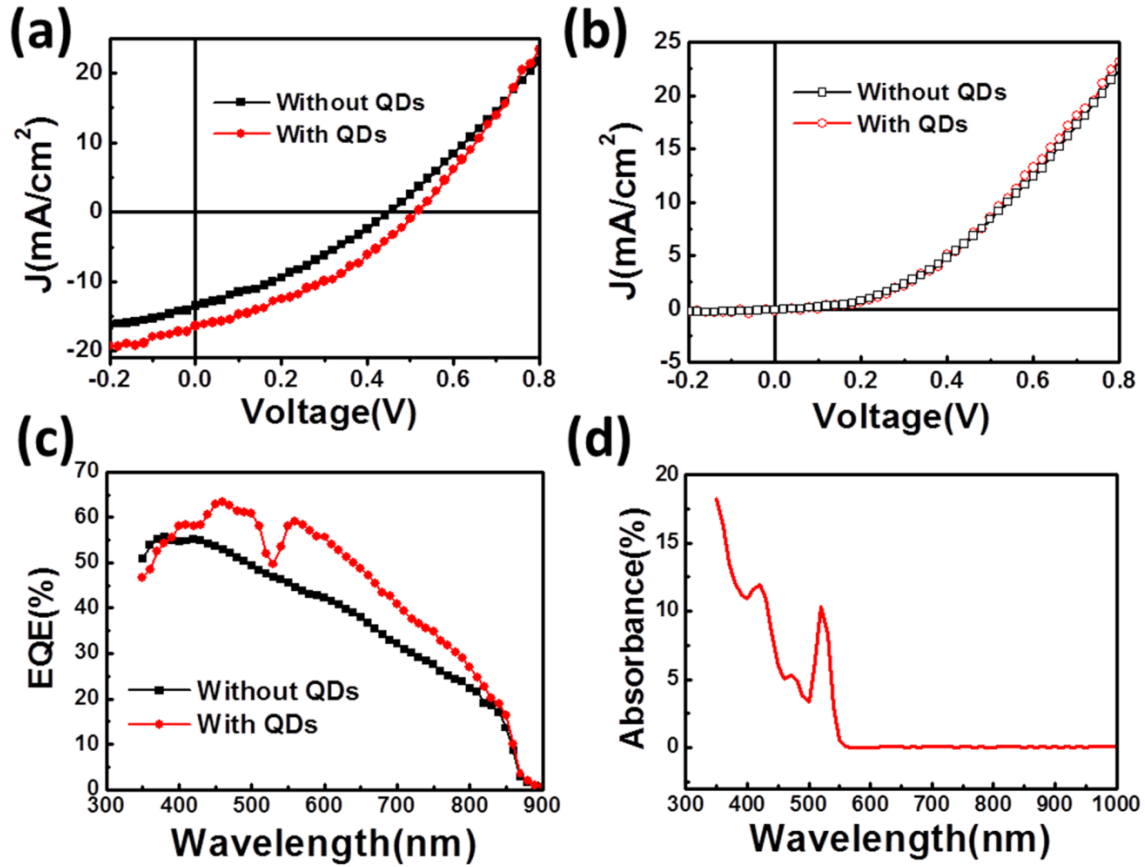

**Fig. 4.** J-V curves under AM1.5G illumination (a) and in the dark (b) of the devices with and without CdSe QDs. (c) EQE curves of the devices with and without CdSe QDs. (d) Absorbance spectrum of CdSe QDs.

Internal quantum efficiency (IQE) curves of the devices with and without QDs in the wavelength range of 450 nm -900 nm are shown in Figure 5(a), where IQE of the device with QDs is higher than that of the device without QDs, which attributes to the photo-induced doping effect. The schematic electronic band structure of the CdSe QDs/graphene/CdTe structure is shown in Figure 5(b). Incident light is mainly absorbed in CdTe layer for the device without QDs. The light induced excess electrons and holes are collected by graphene and CdTe respectively. Short wavelength photons are absorbed near the graphene/CdTe interface, where the generated carriers can be effectively collected. The excess carriers generated by long wavelength photons need to diffuse several micrometers until they are collected, which leads to a lower value of IQE at long wavelengths. For the device with QDs, besides the carriers produced in the CdTe substrate, the excess electrons generated in QDs can hop from QDs directly into graphene. The generation-hopping-collection process between QDs and graphene occurs in atomic scale guaranteeing the high efficiency of this process, thus IQE is increased by introducing photo-induced doping with CdSe QDs. Electron affinity for CdSe, graphene and CdTe are 4.3 eV, 4.6 eV and 4.5 eV respectively. After forming the solar cell, both the valence and conduction bands of both CdSe and CdTe bend downward near the graphene layer. As there are many defects states in the forbidden band gap of CdSe QDs induced by surface dangling bonds, electron-hole pairs can be generated by the incident light in the range of 500 nm-850 nm and the generated electrons are injected into graphene. The photo-induced doping can effectively move up the Fermi level of graphene, even causing the transformation of conduction type from p-type to n-type, which enhances the collection of electrons by graphene layer and results in an improvement of $V_{oc}$, $J_{sc}$, IQE and PCE. The PL band of 450 nm - 850 nm of the QDs shown in Figure 2(a) agrees well with the increased value of IQE after coating with QDs as shown in Figure 5(a), which strongly supports that the photo-induced doping is originated from the carrier generation process of CdSe QDs layer under light illumination.

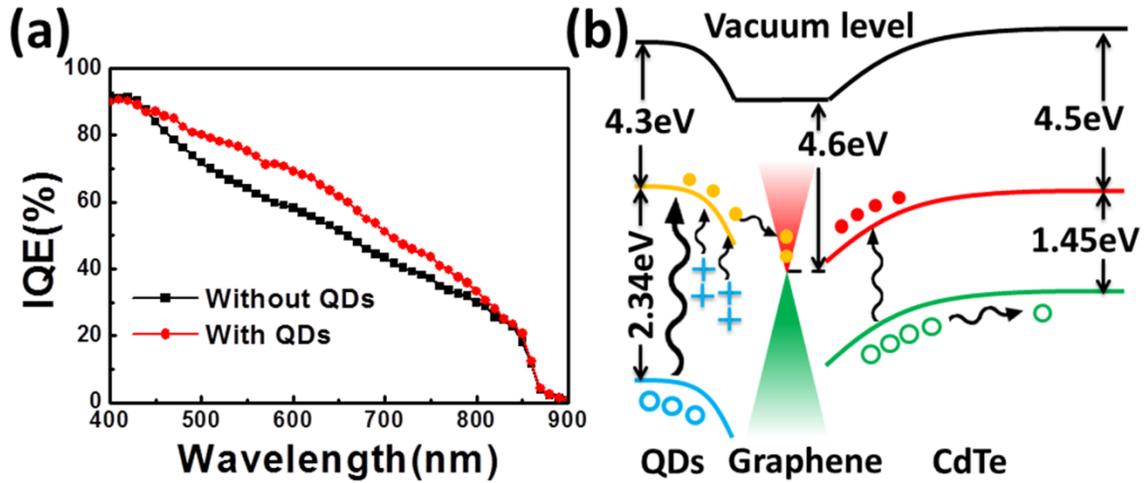

**Fig. 5.** (a) IQE curves of the devices with and without CdSe QDs. (b) Schematic electronic band structure alignment of CdSe QDs covered graphene/CdTe solar cell.

In summary, photo-induced doping enhanced solar cell is reported based on the graphene/CdTe heterostructure. The $J_{sc}$ of the solar cell after coating QDs increases by about 20%, meanwhile PCE increases from original 2.08% to 3.10%. The good match between the PL band from 450 nm - 850 nm and the improved IQE in the range of 450 nm - 850 nm confirm the photo-induced doping effect is accounted for the enhancement of graphene/CdTe solar cells. Considering that the first reported PCE of graphene/Si solar cell is 1.65% in the year 2010, [29] and the PCE value has been improved up to 14.5% in the year of 2015, [30] this first work on graphene/CdTe designed with photo-induced doping holds great promise for large improvements based on interface engineering, process optimization, antireflection coating and so on.

S. S. Lin thanks the support from the National Natural Science Foundation of China (No.51202216, 61431014, 61171037, 61171038, 61322501, 61275183, 61376118, 60990320 and No. 60990322) and Special Foundation of Young Professor of Zhejiang

University (No. 2013QNA5007). X. Q. Li thanks the support from the China Postdoctoral Science Foundation (2013M540491)